\documentclass[aps,prl,twocolumn,showpacs]{revtex4}
\usepackage{graphics}
\usepackage{amsmath}
\usepackage{graphicx}
\usepackage{amsfonts}
\usepackage{amssymb}
\newtheorem{theorem}{Theorem}

\begin{document}
\title{Quantum Reading of a Classical Digital Memory}
\author{Stefano Pirandola}
\affiliation{Department of Computer Science, University of York, York YO10 5GH, United Kingdom}
\date{\today }

\begin{abstract}
We consider a basic model of digital memory where each cell is composed of a
reflecting medium with two possible reflectivities. By fixing the mean number
of photons irradiated over each memory cell, we show that a non-classical
source of light can retrieve more information than any classical source. This
improvement is shown in the regime of few photons and high reflectivities,
where the gain of information can be surprising. As a result, the use of
quantum light can have non-trivial applications in the technology of digital
memories, such as optical disks and barcodes.

\end{abstract}

 \maketitle

In recent years, non-classical states of radiation have been exploited to
achieve marvellous results in quantum information and computation
\cite{Books}. In the language of quantum optics, the bosonic states of the
electromagnetic field are called \textquotedblleft classical\textquotedblright%
\ when they can be expressed as probabilistic mixtures of coherent states.
Classical states describe practically all the radiation sources which are used
in today's technological applications. By contrast, a bosonic state is called
\textquotedblleft non-classical\textquotedblright\ when its decomposition in
coherent states is non-positive \cite{CLASSstate,Prepres}. One of the key
properties which makes a state non-classical is quantum entanglement. In the
bosonic framework, this is usually present under the form of
Einstein-Podolsky-Rosen (EPR) correlations, meaning that the position and
momentum quadrature operators of two bosonic modes are so correlated as to
beat the standard quantum limit \cite{Books}. This is a well-known feature of
the two-mode squeezed vacuum (TMSV) state \cite{Books}, one of the most
important states routinely produced in today's quantum optics labs.

In this Letter, we show how the use of non-classical light possessing
EPR\ correlations can \textit{widely} improve the readout of information from
digital memories. To our knowledge, this is the first study which proves and
quantifies the advantages of using non-classical light for this fundamental
task, being absolutely non-trivial to identify the physical conditions that
can effectively disclose these advantages (as an example, see the recent no-go
theorems of Ref.~\cite{Nair} applied to quantum illumination \cite{QiLL}). Our
model of digital memory is simple but can potentially be extended to realistic
optical disks, like CDs and DVDs, or other kinds of memories such as barcodes.
In fact, we consider a memory where each cell is composed of a reflecting
medium with two possible reflectivities, $r_{0}$ and $r_{1}$, used to store a
bit of information. This memory is irradiated by a source of light which is
able to resolve every single cell. The light focussed on, and reflected from,
a single\ cell is then measured by a detector, whose outcome provides the
value of the bit stored in that cell. Besides the \textquotedblleft
signal\textquotedblright\ modes irradiating the target cell, we also consider
the possible presence of ancillary \textquotedblleft idler\textquotedblright%
\ modes which are directly sent to the detector. The general aim of these
modes is to improve the performance of the output measurement by exploiting
possible correlations with the signals. Adopting this model and fixing the
mean number of photons irradiated over each memory cell, we show that a
\textit{non-classical} source of light with EPR correlations between signals
and idlers can retrieve more information than any classical source of light.
In particular, this is proven for high reflectivities (typical of optical
disks) and few photons irradiated. In this regime the difference of
information can be surprising, up to 1 bit per cell (corresponding to the
extreme situation where only quantum light can retrieve information).
As we will discuss in the conclusion, the chance of reading
information using few photons can have remarkable consequences in
the technology of digital memories, e.g., in terms of
data-transfer rates and storage capacities.\begin{figure}[ptbh]
\vspace{-0.4cm}
\par
\begin{center}
\includegraphics[width=0.5\textwidth] {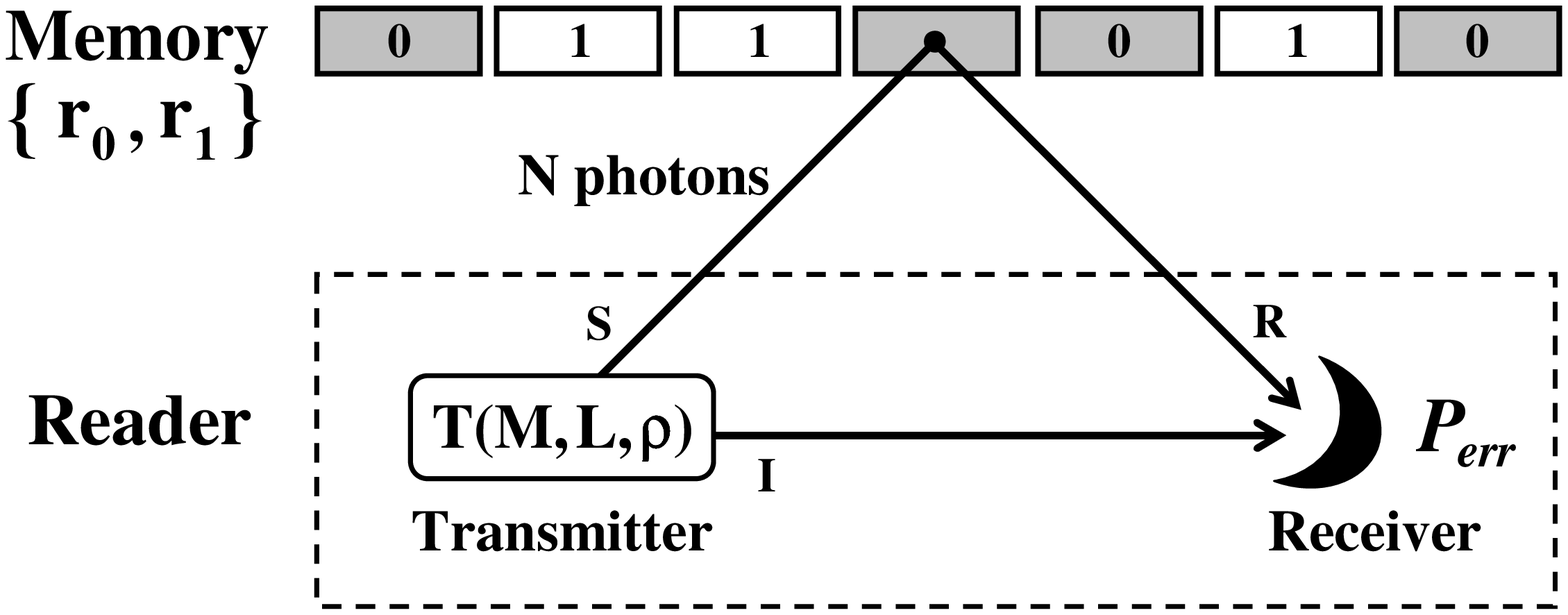}
\end{center}
\par
\vspace{-1.9cm}\caption{\textbf{Basic model of memory}. Digital
information is stored in a memory whose cells have different
reflectivities: $r=r_{0}$ encoding bit-value $u=0$, and $r=r_{1}$
encoding bit-value $u=1$. \textbf{Readout of the memory}. In
general, a digital reader consists of transmitter and receiver.
The transmitter $T(M,L,\rho)$ is a bipartite bosonic system,
composed by a signal system $S$ (with $M$ modes) and an idler
system $I$ (with $L$ modes), which is given in some global state
$\rho$. The signal
$S$ emitted by this source has \textquotedblleft bandwidth\textquotedblright%
\ $M$\ and \textquotedblleft energy\textquotedblright\ $N$\ (mean number of
photons). The signal is directly shined over the cell, and its reflection $R$
is detected together with the idler $I$ at the output receiver, where a
suitable measurement retrieves the value of the bit up to an error probability
$P_{err}$.}%
\label{QreadPIC}%
\end{figure}

Let us consider a digital memory where each cell can have two possible
reflectivities, $r_{0}$ or $r_{1}$, encoding the two values of a logical bit
$u$ (see\ Fig.~\ref{QreadPIC}). Close to the memory, we have a digital reader,
made up of transmitter and receiver, whose goal is to retrieve the value of
the bit stored in a target cell. In general, we call the \textquotedblleft
transmitter\textquotedblright\ a bipartite bosonic system, composed by a
signal system $S$ with $M$ modes and an idler system $I$ with $L$ modes, and
globally given in some state $\rho$. This source can be completely specified
by the notation $T(M,L,\rho)$. By definition, we say that the transmitter $T$
is \textquotedblleft classical\textquotedblright\ (\textquotedblleft
non-classical\textquotedblright) when the corresponding state $\rho$ is
classical (non-classical), i.e., $T_{c}=T(M,L,\rho_{c})$\ and $T_{nc}%
=T(M,L,\rho_{nc})$. The signal $S$ emitted by the transmitter is associated
with two basic parameters: the number of modes $M$, that we call the
\textquotedblleft bandwidth\textquotedblright\ of the signal, and the mean
number of photons $N$, that we call the \textquotedblleft
energy\textquotedblright\ of the signal \cite{NOTEonN}. The signal $S$ is
shined directly on the target cell, and its reflection $R$ is detected
together with the idler $I$ at the output receiver. Here a suitable
measurement yields the value of the bit up to an error probability $P_{err}$.
Repeating the process for each cell of the memory, the reader retrieves an
average of $1-H(P_{err})$\ bits per cell, where $H(\cdot)$ is the binary
Shannon entropy.

The basic mechanism in our model of digital readout is quantum channel
discrimination. In fact, encoding a logical bit $u\in\{0,1\}$\ in a pair of
reflectivities $\{r_{0},r_{1}\}$ is equivalent to encoding $u$ in a pair of
attenuator channels $\{\mathcal{E}(r_{0}),\mathcal{E}(r_{1})\}$, with linear
losses $\{r_{0},r_{1}\}$ acting on the signal modes. The readout of the bit
consists in the statistical discrimination between $r_{0}$\ and $r_{1}$, which
is formally equivalent to the channel discrimination between $\mathcal{E}%
(r_{0})$ and $\mathcal{E}(r_{1})$. The error probability affecting the
discrimination $\mathcal{E}(r_{0})\neq\mathcal{E}(r_{1})$\ depends on both
transmitter and receiver. For a \textit{fixed} transmitter $T(M,L,\rho)$, the
pair $\{\mathcal{E}(r_{0}),\mathcal{E}(r_{1})\}$\ generates two possible
output states at the receiver, $\sigma_{0}(T)$ and $\sigma_{1}(T)$. These are
expressed by $\sigma_{u}(T)=[\mathcal{E}(r_{u})^{\otimes M}\otimes
\mathcal{I}^{\otimes L}](\rho)$, where $\mathcal{E}(r_{u})$ acts on the
signals and the identity $\mathcal{I}$ on the idlers. By optimizing over the
output measurements, the minimum error probability which is achievable by the
transmitter $T$ in the channel discrimination $\mathcal{E}(r_{0}%
)\neq\mathcal{E}(r_{1})$\ is equal to $P_{err}(T)=(1-D)/2$, where $D$ is the
trace distance between $\sigma_{0}(T)$ and $\sigma_{1}(T)$. Now the crucial
point is the minimization of $P_{err}(T)$ over the transmitters $T$. Clearly,
this optimization must be constrained by fixing basic parameters of the
signal. Here we consider the most general situation where only the signal
energy $N$ is fixed. Under this energy constraint the optimal transmitter $T$
which minimizes $P_{err}(T)$ is unknown. For this reason, it is non-trivial to
ask the following question: does a non-classical transmitter which outperforms
any classical one exist? In other words: given two reflectivities
$\{r_{0},r_{1}\}$, i.e., two attenuator channels $\{\mathcal{E}(r_{0}%
),\mathcal{E}(r_{1})\}$, and a fixed value $N$\ of the signal energy, can we
find any $T_{nc}$ such that $P_{err}(T_{nc})<P_{err}(T_{c})$ for every $T_{c}%
$? In the following we reply to this basic question, characterizing the
regimes where the answer is positive. The first step in our derivation is
providing a bound which is valid for every classical transmitter (see
Ref.~\cite{EPAPS}\ for the proof).

\begin{theorem}
[\textbf{classical discrimination bound}]\label{THEOmain}\textit{Let us
consider the discrimination of two reflectivities }$\{r_{0},r_{1}%
\}$\textit{\ using a classical transmitter }$T_{c}$\textit{\ which signals
}$N$\textit{\ photons. The corresponding error probability satisfies}%
\begin{equation}
P_{err}(T_{c})\geq\mathcal{C}(N,r_{0},r_{1}):=\frac{1-\sqrt{1-e^{-N(\sqrt
{r_{1}}-\sqrt{r_{0}})^{2}}}}{2}~. \label{CB_cread}%
\end{equation}

\end{theorem}

\noindent According to this theorem, all the classical transmitters $T_{c}$
irradiating $N$ photons on a memory with reflectivities $\{r_{0},r_{1}\}$
cannot beat the classical discrimination bound $\mathcal{C}(N,r_{0},r_{1})$,
i.e., they cannot retrieve more than $1-H(\mathcal{C})$ bits per cell.
Clearly, the next step is constructing a non-classical transmitter which can
violate this bound. A possible design is the \textquotedblleft
EPR\ transmitter\textquotedblright, composed by $M$ signals and $M$\ idlers,
that are entangled pairwise via two-mode squeezing. This transmitter has the
form $T_{epr}=T(M,M,\left\vert \xi\right\rangle \left\langle \xi\right\vert
^{\otimes M})$, where $\left\vert \xi\right\rangle \left\langle \xi\right\vert
$ is a TMSV state entangling signal mode $s\in S$ with idler mode $i\in I$. In
the number-ket representation $\left\vert \xi\right\rangle =(\cosh\xi
)^{-1}\sum_{n=0}^{\infty}(\tanh\xi)^{n}\left\vert n\right\rangle
_{s}\left\vert n\right\rangle _{i}$, where the squeezing parameter $\xi$
quantifies the signal-idler entanglement. An arbitrary EPR\ transmitter,
composed by $M$ copies of $\left\vert \xi\right\rangle \left\langle
\xi\right\vert $, irradiates a signal with bandwidth $M$ and energy
$N=M\sinh^{2}\xi$. As a result, this transmitter can be completely
characterized by the basic parameters of the emitted signal, i.e., we can set
$T_{epr}=T_{M,N}$. Then, let us consider the discrimination of two
reflectivities $\{r_{0},r_{1}\}$\ using an EPR transmitter $T_{M,N}$ which
signals $N$ photons. The corresponding error probability is upper-bounded by
the quantum Chernoff bound \cite{QCbound}%
\begin{equation}
P_{err}(T_{M,N})\leq\mathcal{Q}(M,N,r_{0},r_{1}):=\frac{1}{2}\left[
\inf_{t\in(0,1)}\mathrm{Tr}(\theta_{0}^{t}\theta_{1}^{1-t})\right]  ^{M},
\label{QCB_qread}%
\end{equation}
where $\theta_{u}:=[\mathcal{E}(r_{u})\otimes\mathcal{I}](\left\vert
\xi\right\rangle \left\langle \xi\right\vert )$. In other words, at least
$1-H(\mathcal{Q})$ bits per cell can be retrieved from the memory. Exploiting
Eqs.~(\ref{CB_cread}) and~(\ref{QCB_qread}), our main question simplifies to
finding $\bar{M}$\ such that $\mathcal{Q}(\bar{M},N,r_{0},r_{1})<\mathcal{C}%
(N,r_{0},r_{1})$. In fact, this implies $P_{err}(T_{\bar{M},N})<\mathcal{C}%
(N,r_{0},r_{1})$, i.e., the existence of an EPR\ transmitter $T_{\bar{M},N}$
able to outperform any classical transmitter $T_{c}$. This is the result of
the following theorem (see Ref.~\cite{EPAPS} for the proof).

\begin{theorem}
[threshold energy]\textit{For every pair of reflectivities }$\{r_{0},r_{1}%
\}$\textit{\ with }$r_{0}\neq r_{1}$\textit{, and signal energy }%
\begin{equation}
N>N_{th}(r_{0},r_{1}):=\frac{2\ln2}{2-r_{0}-r_{1}-2\sqrt{(1-r_{0})(1-r_{1})}%
}~,
\end{equation}
\textit{there is an }$\bar{M}$\textit{\ such that }$P_{err}(T_{\bar{M}%
,N})<\mathcal{C}(N,r_{0},r_{1})$.
\end{theorem}

\begin{figure}[ptbh]
\vspace{-0.14cm}
\par
\begin{center}
\includegraphics[width=0.47\textwidth] {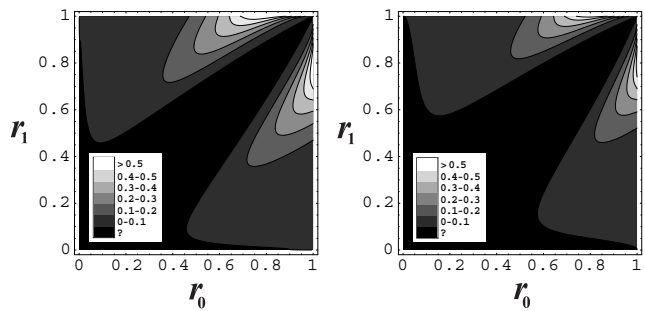}
\end{center}
\par
\vspace{-0.55cm}\caption{\textbf{Left.} Minimum information gain
$G$ over the memory plane $\{r_{0},r_{1}\}$. For a few-photon
signal ($N=30$), we compare a narrowband EPR transmitter ($M=30$)
with all the classical transmitters. Inside the black region
($r_{0}\approx r_{1}$) our investigation is inconclusive. Outside
the black region, we have $G>0$. \textbf{Right.} $G$ plotted over
the plane $\{r_{0},r_{1}\}$ in the presence of decoherence
($\varepsilon=\bar{n}=10^{-5}$). For a few-photon signal ($N=30$),
we compare a narrowband EPR transmitter ($M=30$) with all the
classical transmitters
$T(M,L,\rho_{c})$\ having $M\leq M^{\ast}=5\times10^{6}$.}%
\label{colorbar}%
\end{figure}

\noindent Thus we get the central result of the paper: for every memory and
above a threshold energy, there is an EPR transmitter which outperforms any
classical transmitter. Remarkably, the threshold energy $N_{th}$ turns out to
be low ($<10^{2}$) for most of the memories $\{r_{0},r_{1}\}$ outside the
region $r_{0}\approx r_{1}$. This means that we can have an enhancement in the
regime of few photons ($N<10^{2}$). Furthermore, for low energy $N$, the
critical bandwidth $\bar{M}$\ can be low too. In other words, in the regime of
few photons, narrowband EPR transmitters are generally sufficient to overcome
every classical transmitter. To confirm and quantify this analysis, we
introduce the \textquotedblleft minimum information gain\textquotedblright%
\ $G(M,N,r_{0},r_{1}):=1-H(\mathcal{Q})-[1-H(\mathcal{C})]$. For given memory
$\{r_{0},r_{1}\}$ and signal energy $N$, this quantity lowerbounds the number
of bits per cell which are gained by an EPR\ transmitter $T_{M,N}$ over any
classical transmitter $T_{c}$ \cite{G}. Numerical investigations (see
Fig.~\ref{colorbar}) show that narrowband EPR transmitters are able to give
$G>0$ in the regime of few photons and high reflectivities, corresponding to
having $r_{0}$ or $r_{1}$ sufficiently close to\ $1$ (as typical of optical
disks). In this regime, part of the memories display remarkable gains ($G>0.5$).

Thus the enhancement provided by quantum light can be dramatic in the regime
of few photons and high reflectivities. To investigate more closely this
regime, we consider the case of ideal memories, defined by $r_{0}<r_{1}=1$. As
an analytical result, we have the following \cite{EPAPS}.

\begin{theorem}
[ideal memory]\textit{For every }$r_{0}<r_{1}=1$\textit{\ and
}$N\geq N_{th}=1/2$\textit{, there is a minimum bandwidth
}$\bar{M}$\textit{\ such that
}$P_{err}(T_{M,N})<\mathcal{C}(N,r_{0},r_{1})$\textit{\ for every
}$M>\bar{M}$.
\end{theorem}

\begin{figure}[ptbh]
\vspace{-0.13cm}
\par
\begin{center}
\includegraphics[width=0.47\textwidth] {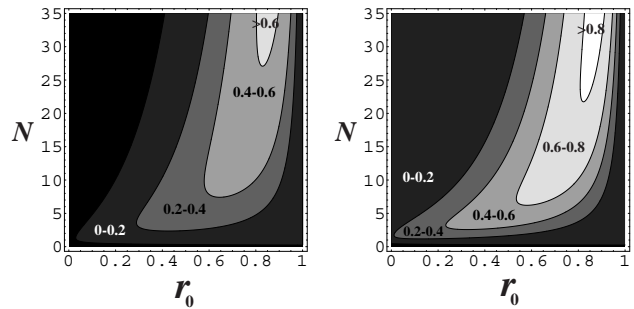}
\end{center}
\par
\vspace{-0.59cm}\caption{Minimum information gain $G$\ versus
$r_{0}$ and $N$. Left picture refers to $M=1$, right picture to
$M\rightarrow\infty$. (For arbitrary $M$ the scenario is
intermediate.) Outside the inconclusive black region we have
$G>0$. For $M\rightarrow\infty$ the black region is completely
collapsed below $N_{th}=1/2$.}%
\label{idealPIC}%
\end{figure}

\noindent Thus, for ideal memories and signals above $N_{th}=1/2$
photon, there are infinitely many EPR transmitters able to
outperform every classical transmitter. For these memories, the
threshold energy is so low that the regime of few photons can be
fully explored. The gain $G$\ increases with the bandwidth, so
that optimal performances are reached by broadband EPR\
transmitters ($M\rightarrow\infty$). However, narrowband EPR
transmitters are sufficient to give remarkable advantages, even
for $M=1$ (i.e., using a single TMSV state). This is shown in
Fig.~\ref{idealPIC}, where $G$ is plotted in terms of $r_{0}$ and
$N$, considering the two extreme cases $M=1$ and
$M\rightarrow\infty$. According to Fig.~\ref{idealPIC}, the value
of $G$ can approach $1$ for ideal memories and few photons even if
we consider narrowband EPR transmitters. 

\textit{Presence of decoherence}.~Note that the previous analysis does not
consider the presence of thermal noise. Actually this is a good approximation
in the optical range, where the number of thermal background photons is around
$10^{-26}$ at about $1~\mu$m and $300~\mathrm{K}$. However, to complete the
analysis, we now show that the quantum effect exists even in the presence of
stray photons hitting the upper side of the memory and decoherence within the
reader. The scattering is modelled as white thermal noise with $\bar{n}$
photons per mode entering each memory cell. Numerically we consider $\bar
{n}=10^{-5}$ corresponding to non-trivial diffusion. This scenario may occur
when the light, transmitted through the cells, is not readily absorbed by the
drive (e.g., using a bucket detector just above the memory) but travels for a
while diffusing photons which hit neighboring cells. Assuming the presence of
one photon per mode travelling the \textquotedblleft optimistic
distance\textquotedblright\ of one meter and undergoing Rayleigh scattering,
we get roughly $\bar{n}\simeq10^{-5}$ \cite{Rayleigh}. The internal
decoherence is modelled as a thermal channel\ $\mathcal{N}(\varepsilon)$
adding Gaussian noise of variance $\varepsilon$ to each signal/reflected mode,
and $2\varepsilon$ to the each idler mode (numerically we consider the
non-trivial value $\varepsilon=\bar{n}=10^{-5}$). Now, distinguishing between
two reflectivities $\{r_{0},r_{1}\}$ corresponds to discriminating between two
Gaussian channels $\mathcal{S}_{u}\otimes\mathcal{N}(2\varepsilon)$ for
$u\in\{0,1\}$. Here $\mathcal{S}_{u}:=\mathcal{N}(\varepsilon)\circ
\mathcal{E}(r_{u},\bar{n})\circ\mathcal{N}(\varepsilon)$ acts on each signal
mode, and contains the attenuator channel $\mathcal{E}(r_{u},\bar{n})$ with
conditional loss $r_{u}$\ and thermal noise $\bar{n}$. To solve this scenario
we use Theorem~\ref{THEOmain} with the proviso of generalizing the classical
discrimination bound. In general, we have $\mathcal{C}=(1-\sqrt{1-F^{M}})/2$,
where $F$ is the fidelity between $\mathcal{S}_{0}(|\sqrt{n_{S}}\rangle
\langle\sqrt{n_{S}}|)$ and $\mathcal{S}_{1}(|\sqrt{n_{S}}\rangle\langle
\sqrt{n_{S}}|)$, the two outputs of a single-mode coherent state $|\sqrt
{n_{S}}\rangle$ with $n_{S}:=N/M$ mean photons. Here the expression for
$\mathcal{C}$ depends also on the bandwidth $M$\ of the classical transmitter
$T_{c}=T(M,L,\rho_{c})$. Since $\mathcal{C}$ decreases to zero for
$M\rightarrow\infty$, our quantum-classical comparison is now restricted to
classical transmitters $T(M,L,\rho_{c})$ with $M$ less than a maximal value
$M^{\ast}<\infty$. Remarkably we find that, in the regime of few photons and
high reflectivities, narrowband EPR\ transmitters are able to outperform all
the classical transmitters up to an extremely large bandwidth $M^{\ast}$. This
is confirmed by the numerical results of Fig.~\ref{colorbar}, proving the
robustness of the quantum effect $G>0$ in the presence of decoherence. Note
that we can neglect classical transmitters with extremely large bandwidths
(i.e., with $M>M^{\ast}$) since they are not meaningful for the model. In
fact, in a practical setting, the signal is an optical pulse with carrier
frequency $\nu$ high enough to completely resolve the target cell. This pulse
has frequency bandwidth $w\ll\nu$ and duration $\tau\simeq w^{-1}$. Assuming
an output detector with response time $\delta t\lesssim\tau$ and
\textquotedblleft reading time\textquotedblright\ $t>\tau$, the number of
modes which are excited is roughly $M=wt$. In other words, the bandwidth\ of
the signal $M$ is the product of its frequency bandwidth $w$ and the reading
time of the detector $t$. Now, the limit $M\rightarrow\infty$ corresponds to
$\delta t\rightarrow0$ (infinite detector resolution) or $t\rightarrow\infty
$\ (infinite reading time). As a result, transmitters with too large an $M$
can be discarded.

\begin{figure}[ptbh]
\par
\begin{center}
\includegraphics[width=0.46\textwidth] {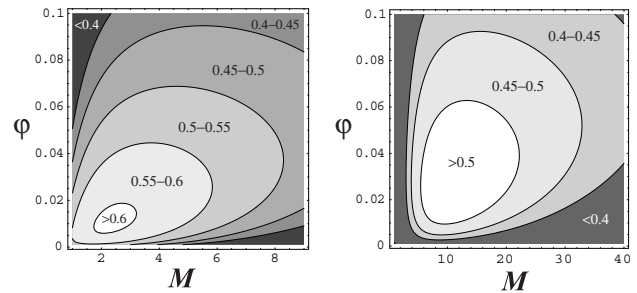}
\end{center}
\par
\vspace{-0.49cm}\caption{\textbf{Left. }$G$ optimized over $M$ and
$\varphi$. $G$ can be higher than $0.6$ bit per cell. Results are
shown in the absence of decoherence ($\varepsilon=\bar{n}=0$)
considering $r_{0}=0.85$, $r_{1}=1$ and $N=35$. \textbf{Right.
}$G$ optimized over $M$ and $\varphi$. Results are shown in the
presence of decoherence ($\bar{n}=\varepsilon=10^{-5}$)
considering $r_{0}=0.85$, $r_{1}=0.95$, $N=100$ and $M^{\ast}=10^{6}$.}%
\label{BellPIC}%
\end{figure}

\textit{Sub-optimal receiver}.~The former results are valid
assuming optimal output detection. Here we show an explicit
receiver design which is (i)\ easy to construct and (ii)\ able to
approximate the optimal results. This sub-optimal receiver
consists of a continuous variable Bell measurement (i.e., a
balanced beam-splitter followed by two homodyne detectors) whose
output is classically processed by a suitable $\chi^{2}$-test with
significance level $\varphi$ (see Ref.~\cite{EPAPS} for details).
In this case the information gain $G$ can be optimized jointly
over the signal bandwidth $M$ (i.e., the number of input TMSV
states)\ and the significance level of the output test $\varphi$.
As shown in Fig.~\ref{BellPIC}, the advantages of quantum reading
are fully preserved.

\textit{Error correction}.~In our basic model of memory we store
one bit of information per cell. In an alternative model,
information is stored in block of cells by using error correcting
codes, so that the readout of data is practically flawless. In
this configuration, we show that the error correction overhead
which is needed by EPR transmitters can be made very small. By
contrast, classical transmitters are useless since they may
require more than 100 cells for retrieving a single bit of
information in the regime of few photons (see Ref.~\cite{EPAPS}
for details).

\textit{Conclusion}.~Quantum reading is able to work in the regime of few
photons. What does it imply? Using fewer photons means that we can reduce the
reading time of the cell, thus accessing higher data-transfer rates. This is a
theoretical prediction that can be checked with a pilot experiment
\cite{EPAPS}. Alternatively, we can fix the total reading time of the memory
while increasing its storage capacity \cite{EPAPS}. The chance of using few
photons leads to another interesting application: the safe readout of
photodegradable memories, such as dye-based optical disks or photo-sensitive
organic microfilms (e.g., containing confidential information.) Here faint
quantum light can retrieve the data safely, whereas classical light could only
be destructive.
More fundamentally, our results apply to the binary discrimination of
attenuator channels.


\begin{thebibliography}{99}
\bibitem {CLASSstate}For any $K$-mode bosonic state $\rho$, we can write the
$\mathcal{P}$-representation $\rho=\int d^{2K}\boldsymbol{\alpha~}%
\mathcal{P}(\boldsymbol{\alpha})\left\vert \boldsymbol{\alpha}\right\rangle
\left\langle \boldsymbol{\alpha}\right\vert $, where $\mathcal{P}%
(\boldsymbol{\alpha})$\ is normalized to $1$ and $\left\vert
\boldsymbol{\alpha}\right\rangle \left\langle \boldsymbol{\alpha}\right\vert $
is a multimode coherent state. Then, $\rho$ is called \textquotedblleft
classical\textquotedblright\ (\textquotedblleft
non-classical\textquotedblright) if $\mathcal{P}(\boldsymbol{\alpha})$ is
positive (non-positive) \cite{Prepres}. For $\rho$ classical, $\mathcal{P}%
(\boldsymbol{\alpha})$ is a probability distribution (implying separability).
\bibitem {G}$G>0$ is a sufficient condition for the superiority of quantum
reading. $G=1$ corresponds to the singular case where only quantum light can
retrieve information.
\end{thebibliography}
\end{document}